\documentclass[12pt,a4paper]{iopart}
\usepackage{graphicx}
\usepackage{iopams}
\usepackage{braket}
\usepackage{setstack} 
\usepackage[square,sort&compress,numbers]{natbib}
\usepackage{color}
\usepackage{hyperref} 
\usepackage{microtype} 
\hypersetup{colorlinks=true,linktoc=all,linkcolor=blue,breaklinks=true,
citecolor=blue,urlcolor=blue}
\usepackage{epstopdf}

\newcommand{\phdagger}{\vphantom{dagger}}
\newcommand{\rhop}{\dot{\rho}} 
\newcommand{\rhow}{\tilde{\rho}} 
\newcommand{\rhowp}{\dot{\tilde{\rho}}}
\newcommand{\HT}{H_\mathrm{T}} 
\newcommand{\HTw}{\tilde{H}_\mathrm{T}}
\newcommand{\Hw}{\tilde{H}}

\begin{document}
\title{Clocked single-spin source based on a spin-split superconductor}
\author{Niklas Dittmann$^{1,2,3}$, Janine Splettstoesser$^{1}$ and Francesco Giazotto$^{4}$}
\address{$^1$ Department of Microtechnology and Nanoscience (MC2), Chalmers 
University of Technology, SE-41298 G{\"o}teborg, Sweden \\
  $^2$ Institute for Theory of Statistical Physics,
      RWTH Aachen, 52056 Aachen,  Germany\\
      and JARA -- Fundamentals of Future Information Technology\\
    $^3$  Peter-Gr\"unberg Institut and Institute for Advanced Simulation,
Forschungszentrum J\"ulich, D-52425 J\"ulich, Germany\\
   $^4$ NEST Istituto Nanoscienze-CNR and Scuola Normale Superiore, I-56127 Pisa, 
Italy
}

\ead{dittmann@chalmers.se}

\begin{abstract}
We propose an accurate clocked single-spin source for ac-spintronic applications. Our device consists of a superconducting island covered by a ferromagnetic insulator layer 
through which it is coupled to superconducting leads. Single-particle transfer relies on the energy gaps and the island's charging 
energy, and is enabled by a bias and a time-periodic gate voltage. Accurate spin transfer is achieved by the ferromagnetic insulator layer
which polarizes the island, provides spin-selective tunneling barriers and improves the precision by suppressing Andreev reflection. We analyze realistic 
material combinations and experimental requirements which allow for a clocked spin current in the MHz regime.
\end{abstract}

\maketitle

\section*{Introduction}
\addcontentsline{toc}{section}{Introduction}
In recent years single-electron sources in solid-state systems have been 
successfully implemented~\cite{Pekola13}, based on superconducting
turnstiles~\cite{Pekola08}, using time-dependently modulated confined structures 
with a discrete spectrum~\cite{Feve07} as well as dynamical quantum dots, driven 
by gating~\cite{Blumenthal07,Kaestner08} or surface-acoustic 
waves~\cite{McNeil11,Hermelin11}. These new types of current sources are 
promising for metrological purposes, they allow to manipulate single particles 
at high frequencies, and are of great interest for quantum computation schemes 
and for the clocked transfer of fundamental units of quantum information.  

Although a number of relevant applications of spintronic 
devices exists~\cite{Wolf01,Zutic04}, the implementation of spintronics at the single-spin level is still weakly explored. 
Only recently, the transfer of single spins between two quantum dots was experimentally reported with a fidelity of 30\%~\cite{Bertrand15}. 
Previous efforts to realize a cyclic electronic pure-spin current source at the single-spin level, instead of stationary spin sources and spin batteries using
rotating magnetic fields~\cite{Brataas02,Mahfouzi10}, are based on a spin ratchet~\cite{Costache10}. 
To our knowledge, single-spin sources with high accuracy are nonetheless still missing. Yet, their successful implementation offers a realm of opportunities: 
for instance they could be used to emit in a controlled way single
quasiparticles with a defined spin into a superconducting contact; this is of interest for spintronics at the single-particle level~\cite{Linder15,Quay15},
for controlled quantum operations (e.g. on flying (spin)-qubits), and for the fundamental research on single-particle characteristics. Furthermore, a 
clocked spin pump, relating the spin current directly to the driving frequency, would provide a very precise spin-current source.

\begin{figure}[t]
\begin{center}
\includegraphics[width=0.7\columnwidth]{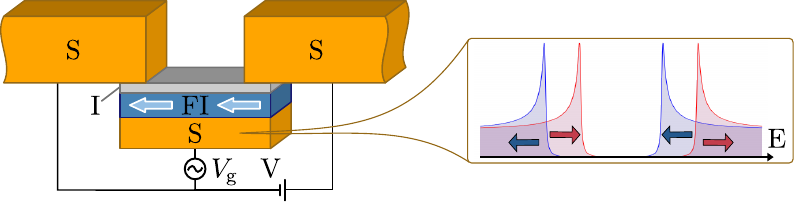}
\end{center}\vspace{-2mm}
	\caption{
Sketch of the turnstile (left); the  ferromagnetic insulator (FI) layer covering the entire thin superconducting (S) island induces a 
spin-split density of states due to exchange interaction (right). Coupling the island to S contacts via the same FI layer provides spin-selective tunneling barriers.
An additional non-magnetic insulator (I) layer prevents a local exchange field in the S contacts \cite{Giazotto15}.}
	\label{fig_setup}
\end{figure}

In this paper, we propose a quantized turnstile acting 
as an accurate \emph{clocked spin source} thanks to the presence of a 
ferromagnetic insulator (FI) layer. As indicated in Figure~\ref{fig_setup}, 
the SFISFIS setup consists of a superconducting (S) island tunnel-coupled to two S leads via a single FI layer.
The island is characterized by a strong charging energy, which together with 
weak tunnel coupling \cite{Giazotto08} implies that the transport of charge and 
spin through the nanostructure takes place by sequential tunneling processes.
As a result of the compact design, the FI layer 
induces a spin-split density of states (DOS) in the small 
island~\cite{Hao90,Meservey94,Moodera07,Li13}, leading to a high spin 
polarization of quasiparticles, and at the same time it provides strongly 
spin-polarized tunneling barriers.\footnote{An alternative setup might use thin FI/I layers at the interfaces between the (serial) S elements
as spin-polarized barriers and an additional thick FI layer on top of the S island to induce the split-field. While this involves changes in the device design,
it does not change the following theoretical investigation.}
At sufficiently low temperatures, the island can be initialized in a state free of quasiparticle excitations~\cite{Maisi13}.
For the turnstile operation, a stationary bias voltage together with a periodically-modulated gate give rise to the generation
of a quasiparticle on the island by an \emph{incoming} charge during the first half of the driving cycle.
Since a single quasiparticle on a superconductor can not relax, which is known as the parity effect \cite{Averin92,Lafarge93,Maisi13,Heimes14},
the island continues to be occupied by one quasiparticle until an annihilation process takes place in the second half of the driving cycle.
This is accompanied by an \emph{outgoing} charge and results in a controlled flow of single particles.
Spin polarization of the generated single-particle current is partially already achieved by the
spin-polarized tunneling barriers. However, here we show that the spin-split DOS of the island is the crucial ingredient for a 
complete spin-polarization of the emitted particles over a wide range of driving frequencies and parameter configurations.
These spin-polarized particles are injected into a nonmagnetic superconducting contact.

In superconductors, as compared to typical semiconducting materials, the quasiparticle spin lifetime 
is largely enhanced~\cite{Meservey83,Yang10,Quay13,Linder15,Eschrig15}. 
This is one of the several reasons why superconducting
spintronics~\cite{Eschrig11,Eschrig15,Linder15} has recently become highly attractive, making the proposed 
spin-turnstile concept very timely.  
Notably, our accurate high-frequency spin source works in the absence of any applied magnetic field. The entirely superconducting 
structure furthermore avoids the technologically difficult combination of superconductors with ferromagnetic metals or halfmetals and is based 
on realizable material combinations and device parameters \cite{Moodera07,Li13,Maisi13}. What is more, an important characteristic of the FI layer 
is that it improves the turnstile precision by strongly suppressing Cooper-pair tunneling and related higher-order processes. 
As a consequence, we expect FI tunnel barriers to be equally beneficial for the precision of pure \emph{charge} turnstiles 
based on superconducting/normal metal (S/N) nanostructures~\cite{Pekola08}, 
which are promising candidates for a new current standard~\cite{Pekola13}.

\section{Superconducting turnstile with ferromagnetic insulator layer}
We describe the superconducting elements of our turnstile by a standard 
Bardeen-Cooper-Schrieffer-Hamiltonian with a (momentum-independent) energy gap $\Delta$. 
The interaction between the localized magnetic moments of the FI  layer and the 
conduction electrons in the superconductor yields an effective exchange-field $h$ in the S island that decays away from the interface
over the superconducting coherence length $\xi_0$~\cite{Tokuyasu88} ($\simeq 100$nm in Al). 
We assume the island thickness to be smaller than $\xi_0$  so that the induced DOS spin-splitting  is
spatially uniform across the entire island~\cite{Hao90,Meservey94,Moodera07,Li13,Giazotto08}.
The system is modeled by the Hamiltonian $H =  H_{\mathrm{contacts}} + H_{\mathrm{island}} + \HT$ with
\begin{eqnarray}
\label{eq_Hamiltonian}
\fl 
H_{\mathrm{contacts}} &= \sum_{a=\mathrm{L,R}} \sum_{\sigma k} \epsilon_{a k}  c_{a \sigma k}^{\dagger} c_{a \sigma k}^{\phdagger} 
     - \sum_{a k} \left( \Delta_{a} c_{a \uparrow k}^{\dagger} c_{a \downarrow -k}^{\dagger} + \Delta_{a}^* c_{a \downarrow -k}^{\phdagger} c_{a \uparrow k}^{\phdagger}\right),
\\  \fl \nonumber
H_{\mathrm{island}} &= \sum_{\sigma k} \left(\epsilon_{k} - \sigma h \right)  d_{\sigma k}^{\dagger} d_{\sigma k}^{\phdagger} 
     - \sum_{k} \left( \Delta d_{\uparrow k}^{\dagger} d_{\downarrow -k}^{\dagger} + \Delta^* d_{\downarrow -k}^{\phdagger} d_{\uparrow k}^{\phdagger}\right)
     + E_{\mathrm{c}} \left( \hat{n}-n_{\mathrm{g}}(t) \right)^2 ,
\\ \fl  \nonumber
\HT &= \sum_{a=\mathrm{L,R}} \sum_{\sigma k l} \left( t^{a\sigma}_{kl} c_{a\sigma l}^{\phdagger} d_{\sigma k}^{\dagger}
     + \mathrm{h.c.} \right) ,
\end{eqnarray}
where $d_{\sigma k}^{(\dagger)}$ and $c_{a\sigma k}^{(\dagger)}$ are electron annihilation (creation) operators for the island and
for the contacts, respectively. All energies are defined with respect to a common equilibrium chemical potential $\mu=0$. 
The subscript $\sigma=\,\,\uparrow,\downarrow$  indicates the quasiparticle spin (parallel/antiparallel to the magnetization of the FI layer), and takes the values $\pm 1$ when used as a variable. 
The island features a strong charging energy, characterized by $E_\mathrm{c}=e^2/(2 C_\Sigma)$ with overall capacitance $C_\Sigma$, where the electron charge is $-e$. 
The charging energy depends on the number of excess charges on the island $n$ (accounted for by the operator $\hat{n}=\sum_{\sigma k}d^\dagger_{\sigma k}d^{\phdagger}_{\sigma k}-n_0$, with offset charge number $n_0$)
with respect to the induced offset charge number $n_\mathrm{g}=C_\mathrm{g}V_\mathrm{g}/e$, where 
$C_\mathrm{g}$ is the gate capacitance and $V_\mathrm{g}$ the gate voltage. 
The Hamiltonian in Equation~(\ref{eq_Hamiltonian}) is diagonalized by a standard Bogoliubov transformation, leading to a description of
the system's excitations in terms of quasiparticles. This and further technical details are presented in Appendix~A.
As a result, the dimensionless quasiparticle DOS of the island, as sketched in Figure~\ref{fig_setup}, can be 
written as
\begin{equation}
g_{\sigma}(E) = \frac{\nu_{\sigma}(E)}{\nu_0} = \left|\mathrm{Re}\,
\left[\frac{E+\sigma h+i\gamma}{\sqrt{(E+\sigma h+i\gamma)^2-\Delta^2}}
\right]\right| ,
\label{eq_dos}
\end{equation}
where $\nu_0$ is  the DOS per spin at the Fermi level in the normal state. 
The dimensionless DOS of the left and right contacts, $g_a(E)$ for $a=\mathrm{L,R}$, is obtained by setting $h=0$ in Equation~(\ref{eq_dos}). 
The Dynes parameter $\gamma$~\cite{Dynes84,Pekola04,Saira12}  accounts for finite broadening in the superconductors.\footnote{The energy gap and the 
Dynes parameter of island and contacts are chosen to be equal for simplicity.
The expected differences in a real device do not change the turnstile working principle.}
The tunneling barriers between island and contacts have spin-dependent contact resistances $R_{a\sigma} = 1/(2 \pi |t^{a\sigma}|^2 \nu_0^2 V_a V_\mathrm{I}))$
with the volumes $V_a$ (contact) and $V_\mathrm{I}$ (island) and $t^{a\sigma}_{kl} = t^{a\sigma}$ assumed to be momentum independent. 
Furthermore, we assume $R_{\mathrm{L}\sigma}=R_{\mathrm{R}\sigma}\equiv R_\sigma$ for simplicity.
The barrier polarization is defined as $P=\left(R_{\downarrow}-R_{\uparrow}\right)/\left(R_{\downarrow}+R_{\uparrow}\right)$.

From the experimental point of view, materials 
such as EuO or EuS, which can provide barrier polarizations as 
high as $\sim 98\%$~\cite{Santos08}, in
contact with superconducting aluminum (Al) are suitable candidates
for the implementation of the spin turnstile. Depending on the thickness of the Al layer and the quality
of the interface, the value for $h$ in such FIS
structures ranges from $\sim 0.2 \Delta$ up to $\sim 0.6\Delta$~\cite{Li13,Li13b,Xiong11,Liu13}.
Alternatively, ferromagnetic GdN barriers combined with superconducting NbN 
could be used with the advantage of a higher critical temperature of $\sim 15$ 
K~\cite{Senapati11,Pal14}. 
In all plots shown below we set $P=90\%$ 
and $h\leq0.3\Delta$. Increasing these parameters would even further improve the  turnstile operation. We furthermore assume the Dynes 
parameter to be of the order of $10^{-5}\Delta$ down to $10^{-6}\Delta$. In analogous devices with non-spin-split superconducting elements, 
the Dynes parameter can reach values down to $10^{-7}\Delta$, favored by the opaque tunnel barriers and further improved by appropriately 
curing the electromagnetic-field environment~\cite{Saira12}. Here, we presume that similar values can be obtained in mesoscopic devices with 
spin-split superconductors~\cite{Giazotto_unpublished}.

\begin{figure}  
\includegraphics[width=\columnwidth]{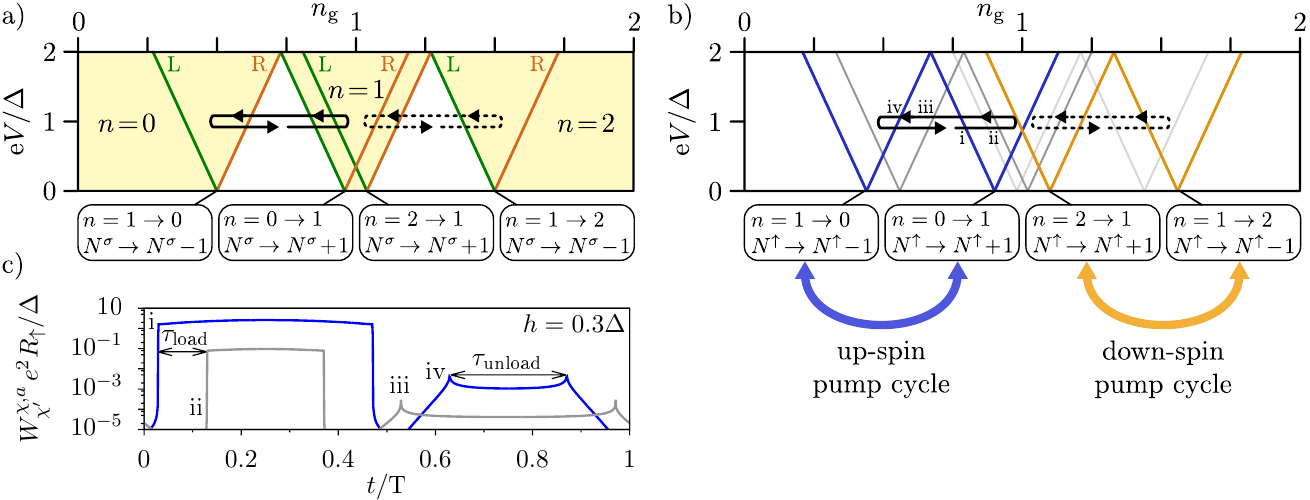}\vspace{-2mm}
	\caption{
a) Stability diagram of a SISIS structure for $E_\mathrm{c}=2.2\Delta$.
Pairs of diagonal lines indicate the set-in of energetically possible tunneling processes: 
Creation/annihilation of island quasiparticles is marked as $N^\sigma \rightarrow N^\sigma\pm1$.
b) Equivalent to a) for a SFISFIS structure with a spin-split island DOS characterized by $h=0.3\Delta$.  Blue/orange lines show processes involving changes in $N^\uparrow$ contributing to pumping of up-spins/down-spins. 
Grey lines indicate changes in $N^\downarrow$, irrelevant for the shown pumping cycles.  
c) Time-evolution of the tunneling rates, where i,ii,iii,iv correspond to crossings of the solid black loop in b) with threshold lines of the same color. 
Further parameters are $eV=\Delta$, $T=0.01T_\mathrm{c}$, $\gamma=10^{-6}\Delta$, $V_\mathrm{I}=1.5\cdot10^{5}\mathrm{nm}^3$, 
$\nu_0=1.45\cdot10^{47}\mathrm{m}^{-3}\mathrm{J}^{-1}$ \cite{Maisi13}, $R_\uparrow=100\mathrm{k}\Omega$, $P=90\%$, with critical temperature $T_\mathrm{c}=1.3\mathrm{K}$ and $\Delta=200\mu \mathrm{eV}$ (aluminum, Al).}
	\label{fig_stationary}  
\end{figure}

\section{Working principle of the clocked spin turnstile}
We now analyze the working principle of the clocked spin turnstile. 
A bias voltage $V$ is symmetrically applied across the structure and the island gate voltage is time-dependently modulated, 
$V_\mathrm{g}(t)=\bar{V}_\mathrm{g}+\delta V_\mathrm{g}A(t)$ and respectively $n_\mathrm{g}(t)=\bar{n}_\mathrm{g}+\delta n_\mathrm{g}A(t)$, where the zero time-average function $A(t)$ 
describes the shape of the driving signal~\cite{Pekola08,Kemppinen09}. This causes tunneling of charges across the device. The addition energies for a charge 
entering $(+)$ or leaving $(-)$ the island, initially occupied with $n$ excess charges, via the left contact are
\begin{eqnarray}
 \label{eq_energydifferences}
 \delta E_+^{\mathrm{L},n} &= E_{\mathrm{c}} \left[(n+1-n_{\mathrm{g}})^2 - (n-n_{\mathrm{g}})^2 \right] - \frac{eV}{2} ,\\ \nonumber
 \delta E_-^{\mathrm{L},n} &= E_{\mathrm{c}} \left[(n-1-n_{\mathrm{g}})^2 - (n-n_{\mathrm{g}})^2 \right] + \frac{eV}{2} ,
\end{eqnarray} 
($V/2$ must be replaced by $-V/2$ for tunnel events via R). 
Charge tunneling goes along with the creation or annihilation of quasiparticles on the island and in the reservoirs. On the island, we have to carefully keep track of the number
of quasiparticles to account for the parity effect.\footnote{To fix the convention,  
we set the state of zero excess charges to be a state with an even number of quasiparticles throughout the whole paper. Then, even/odd charge 
states are always states with an even/odd number of quasiparticles.}  
In contrast, in the large reservoirs the distribution of quasiparticles is well described by a Fermi-function at temperature $T$,
$f(E) = 1/(1+\exp\left(E / k_\mathrm{B} T\right))$. For temperatures of the order of tens of mK, as considered here, the occupation of quasiparticles in the reservoirs is 
strongly suppressed. Hence, a sequential tunnel event that turns an even  island charge state into an odd one necessarily breaks up a Cooper pair in the island or in one of the contacts. 
In order to allow that energetically, the addition energy for adding a quasiparticle to the island has to equal $-2\Delta$. 
However, when the DOS is spin-split as proposed here, see Equation~(\ref{eq_dos}), 
the required energy, $-2 \Delta + \sigma h$, is different for different spin species.
In contrast, when the initial island charge state is odd (namely, occupied by one quasiparticle with spin $\sigma$), a sequential tunnel process that annihilates this quasiparticle
becomes favorable when the addition energy is $\sigma h$ (respectively 0 for the nonmagnetic case).

The turnstile cycle  makes use of the above described tunneling processes. 
This is visualized in the stability diagram for an SISIS charge turnstile in Figure~\ref{fig_stationary}~a), which is shown for a comparison, and
for an SFISFIS spin turnstile in Figure~\ref{fig_stationary}~b). The turnstile cycles are indicated as black loops, the full line showing a cycle involving charge transitions between 0 and 1.
In the first half of this driving cycle, tunneling from the left contact increases the island charge by 1 and a quasiparticle is generated on the island. Due to the presence of the charging energy, further tunneling is suppressed. In the second half of the cycle, one charge leaves the island towards the right lead, while an existing quasiparticle is annihilated. 
Here, we focus on a clocked spin pump with a spin-split island DOS. The onset of a tunneling process therefore depends on the spin of the participating quasiparticle, as shown in Figure~\ref{fig_stationary}~b).
The result is an \textit{up-spin} pump cycle between the charge states $0 \leftrightarrow 1$.  The black dashed loop in Figure~\ref{fig_stationary}~a) and b) shows a second possible driving cycle between
the charge states $1 \leftrightarrow 2$, leading to \textit{down-spin} pumping in the SFISFIS structure. However, we will show that the up-spin pump cycle is favored by the spin-dependent tunnel resistances.

\section{Calculation of the charge and spin current}
For a quantitative analysis of the clocked spin pump, we investigate the probabilities $P(n,N^\uparrow,N^\downarrow)$ that the island holds $n$ excess charges and
$N^{\uparrow}$ and $N^{\downarrow}$ quasiparticle excitations of respective spin. Since the parity of excess charges equals the parity of quasiparticles, the 
occupation probabilities are restricted to the ones, where $n$ and $N^{\uparrow}+N^{\downarrow}$ are both even or both odd. Similarly to previous studies on S/N hybrid structures, 
see e.g.~\cite{Heimes14} and Appendix~C, we derive a Master equation in the sequential tunneling limit, describing the time evolution of the occupation probabilities in the SFISFIS setup,
\begin{eqnarray}
  \label{eq_masterequation}
 \frac{d}{dt} P(\chi) &= \sum_{\chi'} \bigg[ W^{\chi'}_{\chi} P(\chi') - W^{\chi}_{\chi'} P(\chi) \bigg].
\end{eqnarray}
Here, $W^{\chi}_{\chi'} = \sum_{a=\mathrm{L,R}} W^{\chi;a}_{\chi'}$ is a transition rate from $\chi$ to $\chi'$ with $\chi=(n,N^\uparrow,N^\downarrow)$ via quasiparticle 
tunneling between island and contacts. These rates contain the superconducting DOS of both island and contacts and the number of already excited 
quasiparticles on the island via the distribution functions $F_{N^\sigma}$. See, for instance, 
the rate for tunneling of a charge towards the island with simultaneous increase of $N^\uparrow$, 
\begin{eqnarray}
\fl
   \label{eq_example_transitionrate}
    W^{n,N^{\uparrow},N^{\downarrow}}_{n+1,N^{\uparrow}+1,N^{\downarrow}} = \sum_{a=\mathrm{L,R}}
     \frac{1}{R_{\uparrow} e^2 } \int_{0}^{\infty} dE \,  g_{\uparrow}(E)
     \left[1-F_{N^\uparrow}(E)\right]\,g_a(E+\delta E_+^{a,n})\,f(E+\delta E_+^{a,n}) &.
\end{eqnarray}
We model the quasiparticle distribution functions by Fermi functions with an \textit{effective}
temperature $T_{N^\sigma}$~\cite{Maisi13,Heimes14,Giazotto07} (details are shown in Appendix~B). 
This temperature is implicitly set by fixing the island's quasiparticle number 
\begin{eqnarray}
 \label{eq_for_islanddistr}
 N^{\sigma} = 2 \, \nu_0 V_\mathrm{I} \int_{0}^{\infty} dE \, g_{\sigma}(E) F_{N^\sigma}(E),
\end{eqnarray}
with the island's volume $V_\mathrm{I}$. With the help of the transition rates and the occupation probabilities obtained from Equation~(\ref{eq_masterequation}),
the charge ($I^{\mathrm{C}}$) and spin current ($I^{\mathrm{S}}$) through the island can be written as
\begin{eqnarray}
\label{eq_chargecurrent}
I^{\mathrm{C}} &= -\frac{e}{2} \sum_{a=\mathrm{L,R}} \sum_{\sigma}  \sideset{}{'}\sum_{n,N^{\uparrow},N^{\downarrow}} P(n,N^{\uparrow},N^{\downarrow}) 
\\ \nonumber & \quad
\times a \Big(  
     W^{n,N^{\sigma};a}_{n+1,N^{\sigma}+1}
   + W^{n,N^{\bar{\sigma}};a}_{n+1,N^{\bar{\sigma}}-1}
   - W^{n,N^{\sigma};a}_{n-1,N^{\sigma}-1}
   - W^{n,N^{\bar{\sigma}};a}_{n-1,N^{\bar{\sigma}}+1}
   \Big) ,
\\
\label{eq_spincurrent}
I^{\mathrm{S}} &= \frac{\hbar}{4} \sum_{a=\mathrm{L,R}} \sum_{\sigma}  \sideset{}{'}\sum_{n,N^{\uparrow},N^{\downarrow}} P(n,N^{\uparrow},N^{\downarrow}) 
\\ \nonumber & \quad
\times a \sigma \Big(  
     W^{n,N^{\sigma};a}_{n+1,N^{\sigma}+1}
   + W^{n,N^{\bar{\sigma}};a}_{n+1,N^{\bar{\sigma}}-1}
   - W^{n,N^{\sigma};a}_{n-1,N^{\sigma}-1}
   - W^{n,N^{\bar{\sigma}};a}_{n-1,N^{\bar{\sigma}}+1}
   \Big) .
\end{eqnarray}
Here, we introduced the notation $\sum_{n,N^{\uparrow},N^{\downarrow}}'=\sum_{n,N^{\uparrow},N^{\downarrow}\mathrm{,\ with\ }p(n)=p(N^{\uparrow}+N^{\downarrow})}$ and $\bar{\sigma}=-\sigma$.
The index $a$ of all tunnel rates in Equations~(\ref{eq_chargecurrent}) and (\ref{eq_spincurrent}) denotes that these rates are taken only for transfer via the $a$ lead, where $a$
takes the values $\pm1$ for L,R when used as a variable.
Besides that, we abbreviated $W^{n,N^{\uparrow},N^{\downarrow};a}_{n+1,N^{\uparrow}+1,N^{\downarrow}}$ and $W^{n,N^{\uparrow},N^{\downarrow};a}_{n+1,N^{\uparrow},N^{\downarrow}+1}$ by 
$W^{n,N^{\sigma};a}_{n+1,N^{\sigma}+1}$, suppressing the index of the quasiparticle number which remains unchanged in the tunneling process, and similarly also for the other transition rates.
Remarkably, the spin current in Equation~(\ref{eq_spincurrent}) can be interpreted as a sum over spin-polarized charge currents, although 
it is known that in a superconductor the spin current is in general determined by the quasiparticle current \cite{Meservey94}. 
However, owing to the even-odd parity effect on the small island, a change in the number of island charges by $\pm1$ causes a change in the number of quasiparticles; 
therefore in the regime of weak coupling and large charging energy analyzed here, spin-polarized charge currents are a meaningful quantity.
We present technical details about the Master equation and the transition rates in the Appendices~A-C.

\section{Clocked spin-polarized transport}
The evolutions of the relevant transition rates in Figure~\ref{fig_stationary}~c) along the black solid driving cycle in Figure~\ref{fig_stationary}~b) 
illustrate the working principle of the clocked spin pump. The rate for a charge tunneling onto the island by creating an up-spin quasiparticle is largely
increased compared to the one for a down-spin during the time span $\tau_{\mathrm{load}}$. Only when the energy for creating a down-spin quasiparticle on
the island can be brought up, the respective tunnel rate increases. However, the island has already been occupied by an additional charge during $\tau_{\mathrm{load}}$ with a high probability, 
making this rate basically irrelevant. In addition, it remains small due to the strongly spin-polarized tunnel resistances, and can even
be fully suppressed by adjusting the driving cycle such that the crossing at (ii) is avoided. 
In the second half of the driving cycle, the rate for annihilating a down-spin quasiparticle sets in before the corresponding rate for an up-spin quasiparticle 
becomes relevant.  Since, however, no down-spin quasiparticle is occupying the island, also this rate is irrelevant for the turnstile operation. Consequently, during 
the time span $\tau_{\mathrm{unload}}$, an up-spin quasiparticle together with one charge leaves the island.

\begin{figure}
\includegraphics[width=1\columnwidth]{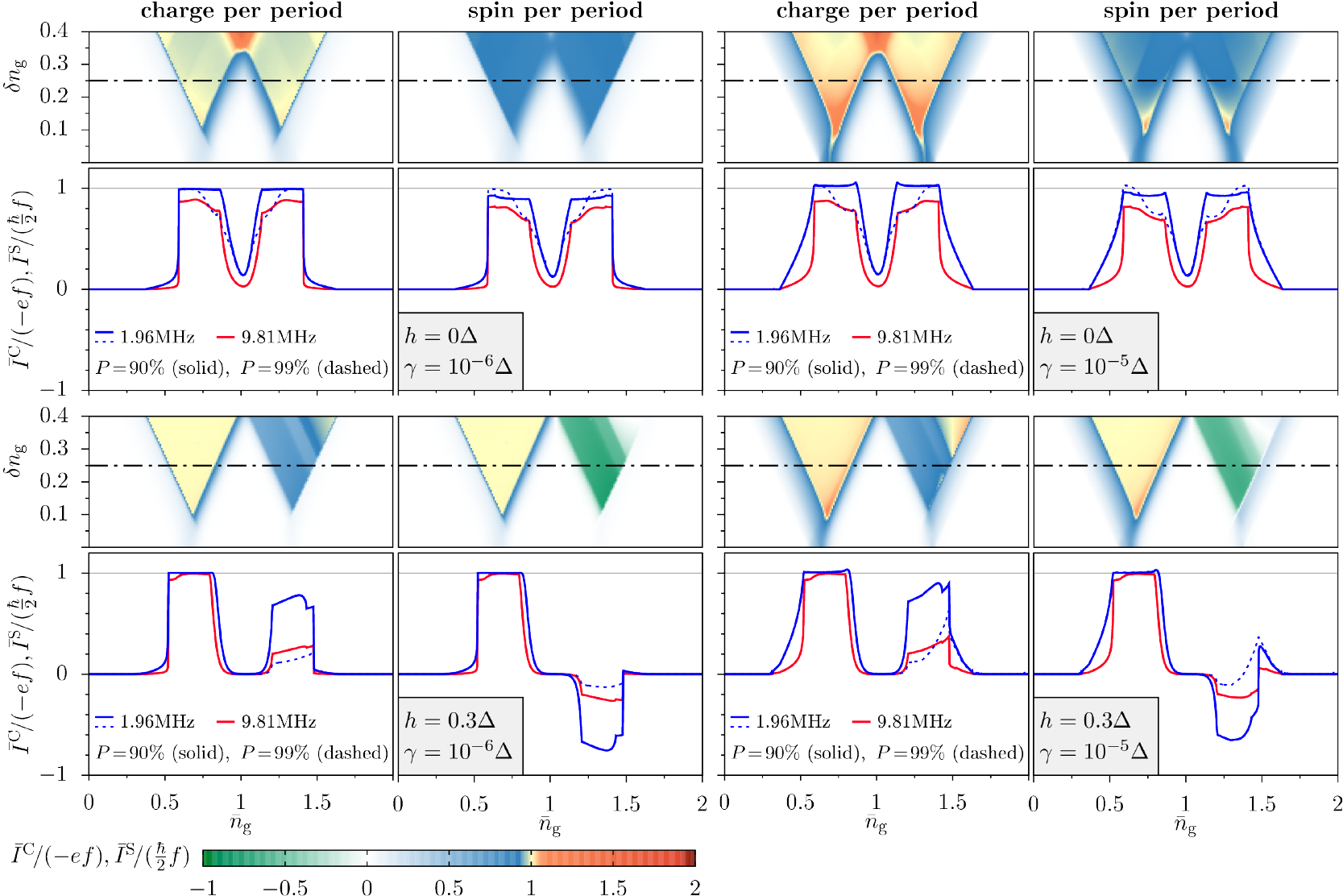}\vspace{-2mm}
\vspace{-2mm}
\caption{
Density plots of the pumped charge ($\bar{I}^C$) and spin ($\bar{I}^S$) per period as a function of the average gate charge 
$\bar{n}_\mathrm{g}$ and the driving amplitude $\delta n_\mathrm{g}$ with and without 
split field $h$ (upper and lower panel) and with $\gamma=10^{-6}\Delta$ and $\gamma=10^{-5}\Delta$ (left and right). Here we set $A(t)=\sin(2\pi f t)$, with driving frequency $f=1.96$MHz, 
$E_\mathrm{c}=2.2\Delta$, $eV=\Delta$, $T=0.01T_\mathrm{c}$, $V_\mathrm{I}=1.5\cdot10^{5}\mathrm{nm}^3$, 
$\nu_0=1.45\cdot10^{47}\mathrm{m}^{-3}\mathrm{J}^{-1}$, $T_\mathrm{c}=1.3\mathrm{K}$, $\Delta=200\mu \mathrm{eV}$, $R_\uparrow=100\mathrm{k}\Omega$ and $P=90\%$.
The dashed-dotted line in each density plot indicates the respective cut which is shown below, for different drive frequencies and barrier polarizations.
}
\label{fig_pump}  
\end{figure}
Figure~\ref{fig_pump} displays the results for the pumped charge ($\bar{I}^C$) and spin ($\bar{I}^S$) per sinusoidal driving cycle as a function of the
working point $\bar{n}_\mathrm{g}$ and the driving amplitude $\delta n_\mathrm{g}$. Let us first analyze the left panels in Figure~\ref{fig_pump}, where the Dynes parameter $\gamma=10^{-6}\Delta$ 
suppresses contributions which arise from the leakage current. Here, the transferred charge is quantized in the expected triangular regions.
The upper row in Figure~\ref{fig_pump} shows that the spin-dependent tunnel resistances already lead to a partially spin-polarized current, even when the split field is neglected. 
However, as clearly visible in the line cuts shown below, the amount of polarization strongly depends on the driving frequency and the chosen working point.
Besides that, the line-cuts for $P=99\%$ show that an increase in the barrier polarization can even be detrimental for the turnstile precision. 
This is a consequence of the spin blockade effect, which occurs if a \emph{down}-spin quasiparticle which has entered the island does not tunnel out during the time $\tau_{\mathrm{unload}}$,
due to fast driving and the reduced tunnel rate for down-spin quasiparticles.
The advantage of a finite  split field $h$ caused by the FI layer is apparent in the second row and the related line cut below: 
the spin-pumping precision is greatly enhanced in the left yellow region. This corresponds to the up-spin pump cycle indicated in Figure~\ref{fig_stationary}~b), 
where the transfer of down-spin particles is energetically blocked. Furthermore, the left plateau in the line-cuts shows that
the precision of this fully spin-polarized clocked current is little sensitive to the driving frequency 
 as well as to small deviations in the working point and the barrier polarization, as 
long as the turnstile operation is enabled. This is in contrast to the green region on the right of the plot, where the turnstile operates as a down-spin pump [see Figure~\ref{fig_stationary}~b)]. 
Here, depending on the spin polarization of the tunnel resistances, the performance is severely limited.

The right panel of Figure~\ref{fig_pump}, compared to the left panel, demonstrates the effect of a larger Dynes parameter $\gamma$ [$10^{-5}\Delta$ instead of $10^{-6}\Delta$]. 
Consequently, the leakage current is enhanced in the right set of plots. A comparison between the left and the right panels reveals regions where the leakage current yields a significant 
contribution to the pumped charge per cycle. This is in particular the case for the extra features occurring at amplitudes $\delta n_\mathrm{g}\lesssim0.1$. Also, 
a smoother transition between regions of vanishing and finite pumped charge and spin can be observed.
In the triangular regions of quantized charge and spin transfer, the enhanced leakage current due to the increased Dynes parameter only leads to slight inaccuracies, 
which can be reduced by increasing the frequency of the periodic driving.
Additional features visible in the density plots in Figure~\ref{fig_pump} are discussed in section~\ref{sec_addfeatures}.

\section{Error sources}
Let us outline possible error sources of the proposed up-spin turnstile. 
A relevant time scale, setting a limit to the operation precision, is given by the inverse of the rate for a charge to tunnel \emph{off} the island by annihilating a quasiparticle. 
During the time $\tau_{\mathrm{unload}}$, indicated in Figure~\ref{fig_stationary}~c), this rate is roughly $W_{\mathrm{unload}} \approx (2 R_\uparrow e^2 \nu_0 V_\mathrm{I})^{-1}\approx 9\mathrm{Mhz}$ 
[for the parameters in Figure~\ref{fig_pump}].
Owing to the large DOS of the island, this rate
is orders of magnitude lower than the rate for tunneling \emph{on} the island with simultaneous creation of a quasiparticle [see Figure~\ref{fig_stationary}~c)]. 
For a precise clocked spin pump, the driving frequency is required to be small enough to provide  $\tau_{\mathrm{unload}} \gtrsim 1/W_{\mathrm{unload}}$. 
However, if the driving frequency is too small, errors might get facilitated due to pair breaking on the island (with a rate of the order of a few kHz \cite{Maisi13}), to spin flips 
(which we here expect to be absent due to the spin-split DOS and the absence of magnetic impurities) and to leakage currents as discussed above
(with rates of the order of $\gamma/(R_\sigma e^2) \approx 10\mathrm{kHz}$ ($100\mathrm{kHz}$) for the parameters in the left (right) panels in Figure~\ref{fig_pump} and $\sigma=\,\,\uparrow$).
Consequently, the described spin-pump operation can only be achieved if $\gamma/(R_\sigma e^2) \ll W_{\mathrm{unload}}$, which restricts
the Dynes parameter to be below $10^{-4}\Delta$ \cite{Giazotto_unpublished}, for all other parameters taken as in Figure~\ref{fig_pump}.
Also, a small driving frequency risks to reduce the magnitude of the spin-polarized current to the noise level of the measurement. This issue can be solved 
by optimizing the driving cycle. The rate $W_{\mathrm{unload}}$ can be increased by decreasing the island size, which we here estimate to be of volume $V_I=200\times 50\times 15\,\mathrm{nm}^3$.  
A way to increase the time  $\tau_{\mathrm{unload}}$ without decreasing the driving frequency is to design the shape of the driving signal appropriately. 
Furthermore, an increase in the generated spin current can also be achieved by operating multiple spin turnstiles in a synchronized way \cite{Maisi09}.

In addition to these limitations, higher-order tunnel processes potentially induce errors, since they enable the tunneling of multiple charges per cycle. 
The dominant processes in second-order tunneling are cotunneling and Andreev reflection. 
However, cotunneling is exponentially suppressed for $eV < 2\Delta$~\cite{Averin08}, as it is the case here. Importantly, the structure proposed here
is well protected against Andreev reflection and other detrimental higher-order processes that involve tunneling of Cooper pairs, in contrast to similar turnstile devices. 
The reason for this are the spin-polarized tunnel barriers which suppress the tunneling of particles with opposite spins. 
For instance, we estimate the tunnel rate for Andreev reflection to be suppressed by a factor of $(1-P)/(1+P)$, when compared to the Andreev tunnel rate through a non-magnetic barrier \cite{Averin08}. 
A detailed analysis of higher-order processes in the presence of spin-polarized barriers and their level of importance is postponed to a future work.

\section{Additional features in the pumped charge and spin}
\label{sec_addfeatures}
\begin{figure}
\includegraphics[width=1\columnwidth]{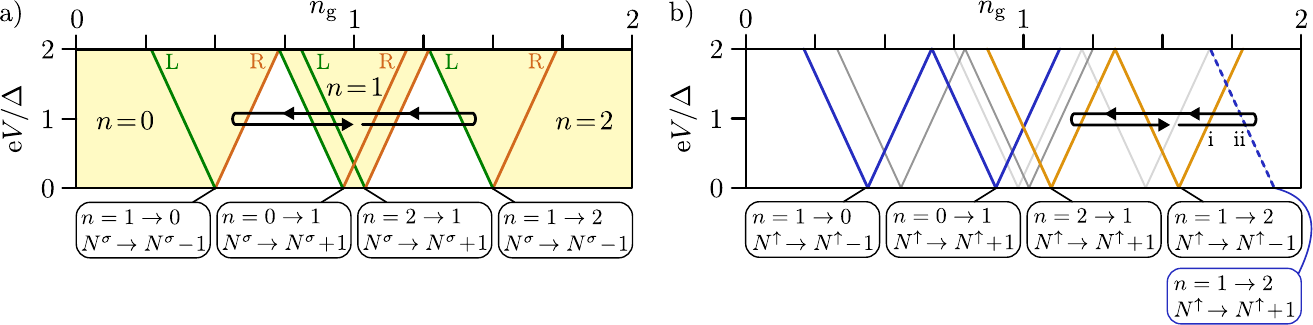}\vspace{-2mm}
		\caption{
a) Similar to Figure~\ref{fig_stationary}~a). The black loop indicates a pump cycle leading to the pumping of two charges per period.
b) Similar to Figure~\ref{fig_stationary}~b). The black loop indicates a pump cycle where 
the transition $n=1\rightarrow2$ via the \emph{creation} of a quasiparticle (blue dashed line) can occur, in the case that
the transition $n=1\rightarrow2$ via the \emph{annihilation} of a quasiparticle did not take place earlier in the pump cycle (e.g.~due to fast driving). 
Crossings with the threshold lines for processes $(n=1\rightarrow2, N^\uparrow\rightarrow N^\uparrow-1)$ via the right contact and $(n=1\rightarrow2, N^\uparrow\rightarrow N^\uparrow+1)$ 
via the left contact are marked by (i) and (ii).
}
	\label{fig_addfeatures}  
\end{figure}

Finally, we discuss additional features visible in the pumped charge and spin per cycle shown in Figure~\ref{fig_pump}, which are however
irrelevant for the proposed spin-turnstile operation.
The first feature which we want to point out occurs in the vicinity of $\bar{n}_\mathrm{g}=1$ for $\delta n_\mathrm{g}>0.3$, as it can be seen in the upper panel for the pumped charge 
for $h=0$. In this region the pumping cycle is large enough to transfer two charges through the device, as indicated in Figure~\ref{fig_addfeatures}~a), 
by combining both driving cycles shown in Figure~\ref{fig_stationary}~b). For the parameters shown,
the transferred charge remains less than 2 since the tunneling of the down-spin is suppressed by the polarized tunnel barriers.

Besides that, two features appear in the upper right part of the density plots for a finite split field (lower panels in Figure~\ref{fig_pump}). 
Towards increasing drive amplitude $\delta n_\mathrm{g}$ and for working points $\bar{n}_\mathrm{g}>1$,
we first find a region where the amount of pumped charge and spin slightly decreases (also visible in the line cuts in the lower panels in Figure~\ref{fig_pump}). 
The decrease is a consequence of the pumping cycle crossing the transition line $(n=1\rightarrow2, N^\uparrow\rightarrow N^\uparrow-1)$ also for the 
$right$ contact. Furthermore, for even larger drive amplitude and for working points $\bar{n}_\mathrm{g}>1$, we find a region where again one charge is pumped per cycle.
The onset of this region coincides with the crossing of the pumping cycle with the transition line $(n=1\rightarrow2, N^\uparrow\rightarrow N^\uparrow+1)$, 
which is shown in Figure~\ref{fig_addfeatures}~b). This means that in this region, the occupation probability $P(2,2,0)$ for two up-spin quasiparticles occupying the island becomes finite and  
contributes to the pumped charge current. Importantly, the pumped spin is suppressed in this region, since both the tunneling of an up-spin and of a down-spin is finite there.
Notably, this process is only possible, if the island can be occupied by two up-spin quasiparticles which do not relax, i.\,e.~recombination processes following a spin flip 
have to be suppressed, as assumed in our model calculation.

\section*{Conclusion}
\addcontentsline{toc}{section}{Conclusion}
In conclusion, we have proposed a clocked, accurate source for single spins based on the parity effect in an S island with S contacts operated as a turnstile. 
The special spin properties of the structure originate from the presence of a ferromagnetic insulator layer which splits the quasiparticle DOS of the island, and at 
the same time provides strongly spin-polarized tunneling barriers, while leaving the contacts nonmagnetic. We emphasize that it is the combination of these effects, 
which provides the possibility of reaching fully polarized, clocked spin currents for spintronic applications over a significant range of driving frequencies and working points.

In addition, we expect these FIS layers to be highly advantageous for \textit{charge} turnstiles, which -- if the precision can be enhanced -- 
are promising candidates for a novel current standard~\cite{Pekola13}.
In these SINIS devices elastic and inelastic cotunneling effects are successfully blocked.
However, they suffer~\cite{Pekola08} from errors deriving from third-order Cooper-pair-electron cotunneling~\cite{Averin08} and 
photon-assisted-tunneling induced Andreev reflection~\cite{diMarco15}. 
We therefore suggest a setup, where the insulating barriers are replaced by FI layers to suppress 
Andreev reflection with the help of strongly spin-polarized barriers. This could greatly 
enhance the precision of clocked charge currents by up to two orders of magnitude for realistic barrier polarizations of $P=98\%$.

\ack 
We thank F.S. Bergeret, A. Di Marco, M. Fogelstr\"om, F. Hassler, 
T. L\"ofwander, J.S. Moodera, D. Persson, P. Samuelsson, and V.S. Shumeiko for helpful discussions. 
We acknowledge funding from the DFG via RTG1995 (N.D., J.S.), from the Knut and 
Alice Wallenberg foundation and the Swedish VR (J.S.),
and the European Research Council under the European Union's Seventh
Framework Programme (FP7/2007-2013)/ERC grant agreement
No. 615187-COMANCHE (F.G.). 

\section*{Appendix}

\subsection*{Appendix A: Charge and quasiparticle degrees of freedom}
\addcontentsline{toc}{section}{Appendix A}
In order to derive a Master equation describing the time evolution of the island's occupation probabilities, 
it is necessary to count the number of excess charges on the 
island. To achieve this, we extend the Hilbert space by charge states $\{\ket{n}\}$~\cite{Schoeller-habil}. 
These states $\{\ket{n}\}$ allow us to keep track of the number of charges that enter or leave the island, without keeping track of their energy distribution.
Note that no coherences between different charge states can occur, since the Josephson energies of the tunnel junctions are small compared to $E_{\mathrm{c}}$.
Starting from the Hamiltonian in Equation~(\ref{eq_Hamiltonian}), the operator $\hat{n}$ in the Hamiltonian is reinterpreted as $\hat{n} \ket{n} = n \ket{n}$ and 
the tunneling part of the Hamiltonian is redefined by adding operators in the $\ket{n}$-subspace, leading to
\begin{eqnarray}
\fl
\label{eq_Hamiltonian_electronoperators2}
H &= \sum_{a=\mathrm{L,R}} \sum_{\sigma k} \epsilon_{a k}  c_{a \sigma k}^{\dagger} c_{a \sigma k}^{\phdagger} 
     - \sum_{a k} \left( \Delta_{a} c_{a \uparrow k}^{\dagger} c_{a \downarrow -k}^{\dagger} + \Delta_{a}^* c_{a \downarrow -k}^{\phdagger} c_{a \uparrow k}^{\phdagger}\right) 
     \\ \fl \nonumber & \quad
     + \sum_{\sigma k} \left(\epsilon_{k} - \sigma h \right)  d_{\sigma k}^{\dagger} d_{\sigma k}^{\phdagger} 
     - \sum_{k} \left( \Delta d_{\uparrow k}^{\dagger} d_{\downarrow -k}^{\dagger} + \Delta^* d_{\downarrow -k}^{\phdagger} d_{\uparrow k}^{\phdagger}\right)
     + E_{\mathrm{c}} \left( \hat{n}-n_{\mathrm{g}}(t) \right)^2
     \\ \fl \nonumber & \quad
     + \sum_{a=\mathrm{L,R}} \sum_{\sigma k l} \,\sum_n \left( t^{a\sigma}_{kl} c_{a\sigma l}^{\phdagger} d_{\sigma k}^{\dagger} \ket{n+1}\!\bra{n}
      + \mathrm{h.c.} \right) .
\end{eqnarray}
The Hamiltonian in Equation~(\ref{eq_Hamiltonian_electronoperators2}) is diagonalized by
applying a Bogoliubov transformation to the electron operators both in the contacts and on the island:
\begin{eqnarray}
 \label{eq_bogoliubovtransformation}
\begin{array}{rlrl}
 \gamma_{\downarrow -k}^{\dagger} &= u_k d_{\downarrow -k}^{\dagger} + v_k^* d_{\uparrow k}^{\phdagger} &\qquad
  \kappa_{a\downarrow -k}^{\dagger} &= u_{ak} c_{a\downarrow -k}^{\dagger} + v_{ak}^* c_{a\uparrow k}^{\phdagger}  \\
 \gamma_{\uparrow k}^{\phdagger} &= u_k^* d_{\uparrow k}^{\phdagger} - v_k d_{\downarrow -k}^{\dagger} &
 \kappa_{a \uparrow k}^{\phdagger} &= u_{ak}^* c_{a\uparrow k}^{\phdagger} - v_{ak} c_{a\downarrow -k}^{\dagger} .
\end{array}
\end{eqnarray}
The result is a description in terms of quasiparticles, where $\gamma_{\sigma k}^{(\dagger)}$ are the island and $\kappa_{a\sigma k}^{(\dagger)}$
the contact quasiparticle operators. The prefactors of the Bogoliubov transformation 
fulfill $|u_k|^2 = \left(1 + \epsilon_k/\sqrt{\epsilon_k^2 + |\Delta|^2}\right)/2$ and
$|v_k|^2 = 1-|u_k|^2$ (equivalently for $|u_{ak}|^2$ and $|v_{ak}|^2$). Using Equation~(\ref{eq_bogoliubovtransformation}), 
the Hamiltonian in Equation~(\ref{eq_Hamiltonian_electronoperators2}) becomes (up to a constant)
\begin{eqnarray}
\fl
H &=  \sum_{a=\mathrm{L,R}} \sum_{\sigma k} E_{a k} \kappa_{a \sigma k}^{\dagger} \kappa_{a \sigma k}^{\phdagger} 
      + \sum_{\sigma k} E_{\sigma k} \gamma_{\sigma k}^{\dagger} \gamma_{\sigma k}^{\phdagger}  + E_{\mathrm{c}} \left( \hat{n}-n_{\mathrm{g}}(t) \right)^2
       + \sum_{a=\mathrm{L,R}} \sum_{k l} \sum_n     
      \\ \fl \nonumber
       & \quad    \Bigg[ t^{a\uparrow}_{kl} \left( u_{al} \kappa_{a \uparrow l}^{\phdagger} + v_{al}\kappa_{a \downarrow -l}^{\dagger}\right)
         \left( u_{k}^* \gamma_{\uparrow k}^{\dagger} + v_{k}^*\gamma_{\downarrow -k}^{\phdagger}\right)\ket{n+1}\!\bra{n}      \\ \fl & \quad \nonumber
       +  t^{a\downarrow}_{kl} 
         \left( u_{al} \kappa_{a \downarrow -l}^{\phdagger} - v_{al}\kappa_{a \uparrow l}^{\dagger}\right)
         \left( u_{k}^* \gamma_{\downarrow -k}^{\dagger} - v_{k}^*\gamma_{\uparrow k}^{\phdagger}\right)\ket{n+1}\!\bra{n}
      + \mathrm{h.c.} \Bigg] ,
\end{eqnarray}
where the quasiparticle energies are $E_{ak} = E(\epsilon_{ak}) = \sqrt{\epsilon_{ak}^2 + |\Delta_{a}|^2}$ for the contacts and 
$E_{\sigma k} = E(\epsilon_{k},\sigma) = -\sigma h + \sqrt{\epsilon_k^2 + |\Delta|^2}$ 
for the Zeeman-split island. 

As pointed out in the main text, the parity effect in the superconducting island~\cite{Averin92,Maisi13,Heimes14} is crucial for the spin-pump operation. To account for the parity effect in our approach,
it is important to keep track of the island quasiparticle excitations of both spin directions by extending the introduced charge states by the quasiparticle numbers $N^\uparrow$ and $N^\downarrow$.
Naturally, the charge number and the quasiparticle numbers are not fully independent of each other: the parity of excess charges $p(n)$ equals the parity 
$p(N^{\uparrow}+N^{\downarrow})$ of the total number of quasiparticle excitations. Therefore, the Hamiltonian is modified to
\begin{eqnarray}
\label{eq_full_ext_Hamiltonian}
\fl 
H &=  \sum_{a=\mathrm{L,R}} \sum_{\sigma k} E_{a k} \kappa_{a \sigma k}^{\dagger} \kappa_{a \sigma k}^{\phdagger} 
      + \sum_{\sigma k} E_{\sigma k} \gamma_{\sigma k}^{\dagger} \gamma_{\sigma k}^{\phdagger}  + E_{\mathrm{c}} \left( \hat{n}-n_{\mathrm{g}}(t) \right)^2
      + \sum_{a=\mathrm{L,R}} \sum_{k l}  \sideset{}{'}\sum_{n,N^{\uparrow},N^{\downarrow}}   \\ \fl \nonumber & \quad
      \Bigg[ t^{a\uparrow}_{kl} 
         \left( u_{al} \kappa_{a \uparrow l}^{\phdagger} + v_{al}\kappa_{a \downarrow -l}^{\dagger}\right)
         \left( u_{k}^* \gamma_{\uparrow k}^{\dagger}\hat{\mathrm{P}}_{n+1,N^\uparrow+1,N^\downarrow}^{n,N^\uparrow,N^\downarrow}
         + v_{k}^*\gamma_{\downarrow -k}^{\phdagger}\hat{\mathrm{P}}_{n+1,N^\uparrow,N^\downarrow-1}^{n,N^\uparrow,N^\downarrow}\right) \\ \fl \nonumber & \quad
      +  t^{a\downarrow}_{kl} 
         \left( u_{al} \kappa_{a \downarrow -l}^{\phdagger} - v_{al}\kappa_{a \uparrow l}^{\dagger}\right)
         \left( u_{k}^* \gamma_{\downarrow -k}^{\dagger}\hat{\mathrm{P}}_{n+1,N^\uparrow,N^\downarrow+1}^{n,N^\uparrow,N^\downarrow} 
         - v_{k}^*\gamma_{\uparrow k}^{\phdagger}\hat{\mathrm{P}}_{n+1,N^\uparrow-1,N^\downarrow}^	{n,N^\uparrow,N^\downarrow}\right) 
      + \mathrm{h.c.} \Bigg] .\nonumber
\end{eqnarray}
Here, we introduced the two abbreviations 
$\hat{\mathrm{P}}^{n',N^{\uparrow'},N^{\downarrow'}}_{n,N^\uparrow,N^\downarrow} = \ket{n,N^\uparrow,N^\downarrow}\!\bra{n',N^{\uparrow'},N^{\downarrow'}}$
and $\sum_{n,N^{\uparrow},N^{\downarrow}}'=\sum_{n,N^{\uparrow},N^{\downarrow}\mathrm{,\ with\ }p(n)=p(N^{\uparrow}+N^{\downarrow})}$.

\subsection*{Appendix B: Density matrix and model for the quasiparticle distribution}
\addcontentsline{toc}{section}{Appendix B}
The added quasiparticle-resolved charge states lead us to a simplified description of the system's dynamics. More precisely, we do not need to take into 
account the exact distribution of quasiparticles and charges over the accessible energy states, when treating the time-evolution of the occupation probabilities, 
$P(n,N^{\uparrow},N^{\downarrow}; t)$, of the island states $|n,N^\uparrow,N^\downarrow\rangle$. In the following, we often suppress 
the time-variable $t$, intending $P(n,N^{\uparrow},N^{\downarrow}; t)\equiv P(n,N^{\uparrow},N^{\downarrow})$.
The density matrix for the extended Hilbert space of the full system is modeled by
\begin{eqnarray}
\fl
 \label{eq_densitymatrix}
 \rho(t) &\approx \rho_{\mathrm{L}}^{\mathrm{eq}} \otimes \rho_{\mathrm{R}}^{\mathrm{eq}} \otimes 
           \sideset{}{'}\sum_{n,N^{\uparrow},N^{\downarrow}} \rho_{\mathrm{island}}^{n,N^{\uparrow},N^{\downarrow}} \otimes 
             \ket{n,N^{\uparrow},N^{\downarrow}}\!\bra{n,N^{\uparrow},N^{\downarrow}} \cdot P(n,N^{\uparrow},N^{\downarrow};t).
\end{eqnarray}
Again, coherent superpositions of island states $|n,N^\uparrow,N^\downarrow\rangle$ with different charge and/or quasiparticle number are not allowed 
due to the island's large charging energy and superconducting gap. In our model, both contacts are assumed to be large reservoirs, which can be described by 
equilibrium density matrices $\rho_{\mathrm{L,R}}^{\mathrm{eq}} $ for all times. Consequently, the quasiparticle distribution function of the contacts 
is given by a Fermi-distribution $f^+(E_{a k}) = (1+e^{E_{a k}/k_{\mathrm{B}} T})^{-1}$ with temperature $T$ (with $f^{-}(E_{ak}) = 1-f^{+}(E_{ak})$).
The density matrix $\rho_{\mathrm{island}}^{n,N^{\uparrow},N^{\downarrow}}$ of the island sub-space is not known in detail, but the separate measurements of 
the excess charge number and the spin-resolved quasiparticle numbers yield $n,N^{\uparrow},N^{\downarrow}$.
The distribution function of island quasiparticles among the quasiparticle energies is defined by
\begin{eqnarray}
 \label{eq_def_distribution_F}
 F^{+}_{n,N^\uparrow,N^\downarrow}(\sigma, k) &= \Tr \left( \gamma_{\sigma k}^{\dagger}\gamma_{\sigma k}^{\phdagger} \,\rho_{\mathrm{island}}^{n,N^{\uparrow},N^{\downarrow}} \right).
\end{eqnarray}
Here, we make the assumptions that $F^{+}_{n,N^\uparrow,N^\downarrow}(\sigma, k)$
only depends on the energy $E_{\sigma k}$ and on the number of quasiparticles with respective spin $N^\sigma$ (thus being independent of the number of excess charges $n$):
\begin{eqnarray}
 F^{+}_{n,N^\uparrow,N^\downarrow}(\sigma, k) &\approx F^{+}_{N^\sigma}(E_{\sigma k}).
\end{eqnarray}
We then model $F^{+}_{N^\sigma}(E_{\sigma k})$ by a Fermi-distribution featuring an effective temperature $T_{N^{\sigma}}$ (with $F^-_{N^\sigma}(E_{\sigma k}) = 1-F^+_{N^\sigma}(E_{\sigma k})$).
The effective temperature, which should not be confused with a physical temperature, is a free parameter which is used
to keep track of the number of quasiparticle excitations on the island. This means that $T_{N^{\sigma}}$ depends on the number of already excited $\sigma$-spin quasiparticles, and is implicitly fixed by the equation
\begin{eqnarray}
  \label{eq_totalFequalsNqp}
  N^\sigma &= \sum_{k} F^+_{N^\sigma}(E_{\sigma k}) =  2 \, \nu_0 V_\mathrm{I} \int_{0}^{\infty} dE \, g_{\sigma}(E) F_{N^\sigma}(E).
\end{eqnarray}
Here, $V_\mathrm{I}$ is the island's volume, $\nu_0$ is the DOS
at the Fermi level in the normal state and $g_{\sigma}(E)$ is the unitless DOS of the superconducting island.
Our model thereby ensures that the occupation numbers of the density matrix in the island quasiparticle subspace are 
in agreement with the number of quasiparticle excitations counted in the additional states $|n,N^\uparrow,N^\downarrow\rangle$. 
In this way, we take care of the parity effect in our model, since the sequential tunnel rates of the Master equation, derived in Appendix~C, explicitly  depend 
on $F^+_{N^\sigma}$ and thus on the number of excited quasiparticles.
In a real system, the distribution of quasiparticles on the island might differ from the Fermi distribution. However, the precise form of
$F^+_{N^\sigma}$ should not influence the working principle of the proposed clocked spin pump, as long as $F^+_{N^\sigma}\ll 1$~\cite{Heimes14}. 

\subsection*{Appendix C: Derivation of the Master equation in Born-Markov approximation}
\addcontentsline{toc}{section}{Appendix C}
We now come to the derivation of the Master equation, which describes the time evolution of the occupation probabilities $P(n,N^{\uparrow},N^{\downarrow})$. 
The Master equation is calculated in Born-Markov approximation, i.\,e.\,\,restricted to sequential tunneling while neglecting memory effects, see for example 
\cite{Schaller14}.
Starting point for our derivation is the Liouville equation
\begin{eqnarray}
  \label{eq_Liouville}
  i \rhop(t) &= \left[ H, \rho(t) \right] .
\end{eqnarray}
For convenience we set $\hbar=1$ and $e=1$ (so that the charge of the electron is $-1$). 
As a first step, we switch to the interaction picture with respect to the perturbation $\HT$, where interaction-picture operators are marked by a $\sim$ symbol. We obtain the
Liouville equation in Born-Markov approximation
\begin{eqnarray}
  \label{eq_Liouville2}
  \rhowp(t) &= (-i)^2 \int_{-\infty}^{t}\! dt'  \left[ \HTw(t), \left[ \HTw(t'), \rhow(t) \right] \right] .
\end{eqnarray}
The time evolution of the probabilities $P(n,N^{\uparrow},N^{\downarrow})$ is calculated by taking the time derivative of the expectation value of the projector 
$\hat{P}_{n,N^{\uparrow},N^{\downarrow}} = \hat{P}_{n,N^{\uparrow},N^{\downarrow}}^{n,N^{\uparrow},N^{\downarrow}} = \ket{n,N^{\uparrow},N^{\downarrow}}\!\bra{n,N^{\uparrow},N^{\downarrow}}$. 
With Equation~(\ref{eq_Liouville2}) follows
\begin{eqnarray}
  \label{eq_Liouville_for_P}
  \dot{P}(n,N^{\uparrow},N^{\downarrow}) &= \Tr\left( \hat{P}_{n,N^{\uparrow},N^{\downarrow}} \, \rhop(t) \right) \\ \nonumber
  &= (-i)^2 \int_{-\infty}^{t}\! dt' \Tr\left( \hat{P}_{n,N^{\uparrow},N^{\downarrow}} \left[ \HTw(t), \left[ \HTw(t'), \rhow(t) \right] \right] \right).
\end{eqnarray}
Before evaluating the double commutator in Equation~(\ref{eq_Liouville_for_P}), we write the tunnel Hamiltonian as 
\begin{eqnarray}
 \HTw(t) &= \sum_{\alpha\beta=\pm} \sum_{\sigma k} e^{i H_0 t} H_{\sigma k}^{\alpha \beta} e^{-i H_0 t}
\end{eqnarray}
where $H_0$ represents the unperturbed Hamiltonian, $\alpha=\pm$ indicates if a charge is added or subtracted from the island and $\beta=\pm$ marks the excitation or annihilation of an island quasiparticle
during the tunnel process $(\bar{\alpha}=-\alpha,\bar{\beta}=-\beta)$. The operators $H_{\sigma k}^{\alpha \beta}$ (with  $H^{\alpha\beta}_{\sigma k} = \left(H^{\bar{\alpha}\bar{\beta}}_{\sigma k}\right)^\dagger$) read
\begin{eqnarray}
 H^{++}_{\uparrow k} &= \sum_{a=\mathrm{L,R}} \,\, \sideset{}{'}\sum_{n,N^{\uparrow},N^{\downarrow}} \sum_{l} t^{a\uparrow}_{kl}
         \left( u_{al} \kappa_{a \uparrow l}^{\phdagger} + v_{al}\kappa_{a \downarrow -l}^{\dagger}\right) u_{k}^* \gamma_{\uparrow k}^{\dagger}\hat{\mathrm{P}}_{n+1,N^\uparrow+1,N^\downarrow}^{n,N^\uparrow,N^\downarrow}  \\ \nonumber
 H^{+-}_{\uparrow k} &= \sum_{a=\mathrm{L,R}}\,\, \sideset{}{'}\sum_{n,N^{\uparrow},N^{\downarrow}} \sum_{l} t^{a\uparrow}_{kl}
         \left( u_{al} \kappa_{a \uparrow l}^{\phdagger} + v_{al}\kappa_{a \downarrow -l}^{\dagger}\right) v_{k}^*\gamma_{\downarrow -k}^{\phdagger}\hat{\mathrm{P}}_{n+1,N^\uparrow,N^\downarrow-1}^{n,N^\uparrow,N^\downarrow}  \\  \nonumber
 H^{++}_{\downarrow k} &= \sum_{a=\mathrm{L,R}} \,\,\sideset{}{'}\sum_{n,N^{\uparrow},N^{\downarrow}} \sum_{l} t^{a\downarrow}_{kl} 
         \left( u_{al} \kappa_{a \downarrow -l}^{\phdagger} - v_{al}\kappa_{a \uparrow l}^{\dagger}\right) u_{k}^* \gamma_{\downarrow -k}^{\dagger}\hat{\mathrm{P}}_{n+1,N^\uparrow,N^\downarrow+1}^{n,N^\uparrow,N^\downarrow} \\ \nonumber
 H^{+-}_{\downarrow k} &= \sum_{a=\mathrm{L,R}} \,\,\sideset{}{'}\sum_{n,N^{\uparrow},N^{\downarrow}} \sum_{l} t^{a\downarrow}_{kl} 
         \left( u_{al} \kappa_{a \downarrow -l}^{\phdagger} - v_{al}\kappa_{a \uparrow l}^{\dagger}\right) (-1) v_{k}^*\gamma_{\uparrow k}^{\phdagger}\hat{\mathrm{P}}_{n+1,N^\uparrow-1,N^\downarrow}^{n,N^\uparrow,N^\downarrow}.
\end{eqnarray}
Plugging these expressions into Equation~(\ref{eq_Liouville_for_P}) and only keeping the parts of the double commutator that do not vanish under the trace leads to
\begin{eqnarray}
\fl
  \label{eq_Liouville_for_P2}
  & \dot{P}(n,N^{\uparrow},N^{\downarrow}) \\ \fl \nonumber
  &= (-i)^2 \int_{-\infty}^{t}\! dt' \Tr\left( \hat{P}_{n,N^{\uparrow},N^{\downarrow}}\sum_{\alpha\beta\sigma k} \left[ \Hw_{\sigma k}^{\alpha\beta}(t), 
        \left[ \Hw_{\sigma k}^{\bar{\alpha}\bar{\beta}}(t'), \rhow(t) \right] \right] \right) \\ \fl \nonumber
  &=  2 \Re \int_{-\infty}^{t}\! dt' \sum_{\alpha\beta\sigma k} \Tr\Bigg( 
         \Hw_{\sigma k}^{\alpha\beta}(t) \hat{P}_{n,N^{\uparrow},N^{\downarrow}} \Hw_{\sigma k}^{\bar{\alpha}\bar{\beta}}(t') \rhow(t)
        - \Hw_{\sigma k}^{\bar{\alpha}\bar{\beta}}(t')  \Hw_{\sigma k}^{\alpha\beta}(t) \hat{P}_{n,N^{\uparrow},N^{\downarrow}} \rhow(t) 
        \Bigg)
        \end{eqnarray}
The 16 objects $ I_{\sigma k}^{\alpha\beta}(t,t') = \Tr \{\Hw_{\sigma k}^{\bar{\alpha}\bar{\beta}}(t')  \Hw_{\sigma k}^{\alpha\beta}(t)\hat{P}_{n,N^{\uparrow},N^{\downarrow}} \rhow(t) \}$ and 
$J_{\sigma k}^{\alpha\beta}(t,t') = \Tr \{\Hw_{\sigma k}^{\alpha\beta}(t) \hat{P}_{n,N^{\uparrow},N^{\downarrow}} \Hw_{\sigma k}^{\bar{\alpha}\bar{\beta}}(t') \rhow(t) \}$ in 
Equation~(\ref{eq_Liouville_for_P2}) are calculated by applying standard manipulations.  We then summarize all terms appearing on the r.\,h.\,s~of Equation~(\ref{eq_Liouville_for_P2}) in products 
containing an occupation probability and a transition rate. This leads to the Master equation for the occupation probabilities
\begin{eqnarray}
\fl
 \label{eq_mastereq}
 \dot{P}(n,N^{\uparrow},N^{\downarrow}) 
 = \sideset{}{''}{\sum}_{n',{N^\uparrow}',{N^\downarrow}'}
   \left[ W^{n',{N^{\uparrow}}',{N^{\downarrow}}'}_{n,N^{\uparrow},N^{\downarrow}} \, P(n',{N^{\uparrow}}',{N^{\downarrow}}')
      -   W^{n,N^{\uparrow},N^{\downarrow}}_{n',{N^{\uparrow}}',{N^{\downarrow}}'} P(n,N^{\uparrow},N^{\downarrow})\right] .
\end{eqnarray}
Here, $\sum''$ is defined as the sum over all combinations of $n',{N^\uparrow}',{N^\downarrow}'$ for which $n'-n=\pm1 $ and ${N^\uparrow}'+{N^\downarrow}'-(N^\uparrow+N^\downarrow)=\pm1$.
In Equation~(\ref{eq_mastereq}), a transition from $(n,N^{\uparrow},N^{\downarrow})$ to $(n',N^{\uparrow'},N^{\downarrow'})$ is characterized by 
the transition rate $W^{n,N^{\uparrow},N^{\downarrow}}_{n',N^{\uparrow'},N^{\downarrow'}}$, where the indices describe the island excess charges 
and quasiparticle excitations before and after the tunnel process. In total, starting from the island's occupation $(n,N^{\uparrow},N^{\downarrow})$, eight different sequential 
tunnel processes can in principle occur. The tunneling of one charge \emph{towards} the island is divided in four different processes:
the involved electron features either an up spin or a down spin and, during the process, a quasiparticle of the same spin is generated 
or a quasiparticle of the opposite spin is annihilated. The tunnel rates of these four processes are
\begin{eqnarray}
\fl
 W^{n,N^{\uparrow},N^{\downarrow}}_{n+1,N^{\uparrow}+1,N^{\downarrow}}
 &= \sum_{a=\mathrm{L,R}} \frac{1}{R_{\uparrow}} \int_{0}^{\infty} dE \, 
     g_a\left(E+\delta E_+^{a,n}\right) g_{\uparrow}(E)\, f^+\!\left(E+\delta E_+^{a,n}\right) F^-_{N^\uparrow}(E)  \\\fl \nonumber
 W^{n,N^{\uparrow},N^{\downarrow}}_{n+1,N^{\uparrow},N^{\downarrow}-1} 
 &= \sum_{a=\mathrm{L,R}} \frac{1}{R_{\uparrow}}  \int_{0}^{\infty} dE \, 
     g_a\left(-E+\delta E_+^{a,n}\right) g_{\downarrow}(E)\, f^+\!\left(-E+\delta E_+^{a,n}\right) F^+_{N^\downarrow}(E)  \\\fl \nonumber
 W^{n,N^{\uparrow},N^{\downarrow}}_{n+1,N^{\uparrow},N^{\downarrow}+1} 
 &= \sum_{a=\mathrm{L,R}} \frac{1}{R_{\downarrow}}  \int_{0}^{\infty} dE \, 
     g_a\left(E+\delta E_+^{a,n}\right) g_{\downarrow}(E)\, f^+\!\left(E+\delta E_+^{a,n}\right) F^-_{N^\downarrow}(E)  \\\fl \nonumber
 W^{n,N^{\uparrow},N^{\downarrow}}_{n+1,N^{\uparrow}-1,N^{\downarrow}}
 &= \sum_{a=\mathrm{L,R}} \frac{1}{R_{\downarrow}} \int_{0}^{\infty} dE \, 
     g_a\left(-E+\delta E_+^{a,n}\right) g_{\uparrow}(E)\, f^+\!\left(-E+\delta E_+^{a,n}\right) F^+_{N^\uparrow}(E) ,
\end{eqnarray}
where we defined the spin-dependent tunnel resistances $R_{a\sigma} = 1/(2 \pi |t^{a\sigma}|^2 \nu_0^2 V_a V_\mathrm{I})) = R_{\sigma}$
with the volumes $V_a$ (contact) and $V_\mathrm{I}$ (island) and $\nu_0$ is the DOS per spin at the Fermi level in the normal state, 
and we assume $t^{a\sigma}_{kl} = t^{a\sigma}$ to be momentum independent.
The dimensionless DOS $g_a(E_a)$ and $g_\sigma(E)$ of the island and the contacts are defined in Equation~(\ref{eq_dos}).
The tunneling of one charge \emph{off} the island is equivalently divided in four processes, where the respective tunnel rates are
\begin{eqnarray}\label{eq_rates2}
\fl
W^{n,N^{\uparrow},N^{\downarrow}}_{n-1,N^{\uparrow},N^{\downarrow}+1}
 &= \sum_{a=\mathrm{L,R}} \frac{1}{R_{\uparrow}} \int_{0}^{\infty} dE \, 
     g_a\left(-E-\delta E_-^{a,n}\right) g_{\downarrow}(E)\, f^-\!\left(-E-\delta E_-^{a,n}\right) F^-_{N^\downarrow}(E)  \\ \fl \nonumber
 W^{n,N^{\uparrow},N^{\downarrow}}_{n-1,N^{\uparrow}-1,N^{\downarrow}}
 &= \sum_{a=\mathrm{L,R}} \frac{1}{R_{\uparrow}} \int_{0}^{\infty} dE \, 
     g_a\left(E-\delta E_-^{a,n}\right) g_{\uparrow}(E)\, f^-\!\left(E-\delta E_-^{a,n}\right) F^+_{N^\uparrow}(E) \\ \fl \nonumber
 W^{n,N^{\uparrow},N^{\downarrow}}_{n-1,N^{\uparrow}+1,N^{\downarrow}}
 &= \sum_{a=\mathrm{L,R}} \frac{1}{R_{\downarrow}}  \int_{0}^{\infty} dE \, 
     g_a\left(-E-\delta E_-^{a,n}\right) g_{\uparrow}(E)\, f^-\!\left(-E-\delta E_-^{a,n}\right) F^-_{N^\uparrow}(E) \\ \fl \nonumber
 W^{n,N^{\uparrow},N^{\downarrow}}_{n-1,N^{\uparrow},N^{\downarrow}-1}
 &= \sum_{a=\mathrm{L,R}} \frac{1}{R_{\downarrow}}  \int_{0}^{\infty} dE \, 
     g_a\left(E-\delta E_-^{a,n}\right) g_{\downarrow}(E)\, f^-\!\left(E-\delta E_-^{a,n}\right) F^+_{N^\downarrow}(E) .
\end{eqnarray}
Each of the eight tunnel processes can only take place if the addition energy $\delta E_\pm^{a,n}$ defined in Equation~(\ref{eq_energydifferences}) is brought up. 
This includes the change of charging energy and a contribution from an applied bias voltage $V$. Notably, the tunnel rates in Equation~(\ref{eq_mastereq})-(\ref{eq_rates2}) depend on time
via the addition energies.

\providecommand{\newblock}{}

\end{document}